# Updated design for LEP3

M. Koratzinos, P. Raimondi, CERN, Geneva, Switzerland

*Abstract*

An updated design for the LEP3 electron–positron collider is presented. The machine is designed to operate in the existing tunnel infrastructure currently hosting the Large Hadron Collider and aims to deliver high luminosity for precision studies of the Z, W, and Higgs boson. The design adopts, wherever possible, technical solutions developed for the Future Circular Collider electron–positron program (FCC-ee) and adapts them to the constraints of the LHC tunnel. The proposed lattice includes a low-emittance arc structure, long straight sections optimized for RF installations, and a final focus system derived from the FCC-ee design. Two experimental insertions are envisaged, at IP1 and IP5 where currently Atlas and CMS reside which could be adapted versions of the current detectors. Hardware requirements, RF system configurations, and booster ring optics are discussed, together with preliminary estimates of performance and operational considerations.

## INTRODUCTION

The LEP3 idea was introduced at the time of the discovery of the Higgs boson at CERN, following a seminal paper by Blondel and Zimmermann [1], see for instance [2]. The idea was recently revived as a low-cost alternative to the FCC-ee project [3]. Since then, we have upgraded significantly the LEP3 baseline with a number of improvements and enhancements, the major one being a new low-emittance lattice that also manages lower energy loss per turn.

We are opting for luminosities of 50, 10 and 3E34 for the Z, W and H running, which we believe can be achieved while keeping below the SR power budget of 50MW.

## FROM LEP AND LEP2 TO LEP3

LEP3 has orders of magnitude higher luminosity than LEP and LEP2. Major improvements come from the following: separate beam pipes for electrons and positrons allow for a much higher number of bunches; this, in conjunction with the higher power provided by the RF power sources, increase beam current and luminosity. The crab waist scheme allows to take full advantage of very small beta* values, increasing luminosity. The use of continuous injection using a separate booster ensures that average luminosity is close to peak luminosity. Finally, lower horizontal and, therefore, vertical emittance helps increase the Piwinski angle and the hourglass factor which increase luminosity and also decreases the bunch population for the same beam-beam parameter, which decreases beamstrahlung.

## LEP3 DESIGN STRATEGY

We sre not starting from scratch in this exercise; a lot of work has already been done in the FCC-ee design. Our aim is to design LEP3 so that we minimise capital expenditure and operational expenditure, achieve maximum reliability with a high confidence level and at the same time maximise performance. Therefore we:

- Adopt beam parameters and solutions already developed for FCC-ee
- Exploit synergies with FCC-ee
- Optimize the ring lattice/layout to relax requirements on beam parameters and hardware
- Eventually we wish to establish a parametric risk analysis that will allow to estimate a confidence level for achieving the required beam parameters and luminosity goals and define a methodology to increase such confidence level.
- Define a upgrade path whenever relevant and a timeline for its implementation

The LEP3 design challenges stem from the fact that we would like to reuse existing infrastructure wherever possible. This imposes limitations on:

- ring circumference,
- straight section locations,
- interaction region geometry.

This is especially challenging when designing the final focus system. We are optimizing hardware for cost, power consumption and reliability.

## ARC LATTICE DESIGN

The arc lattice is designed to produce horizontal emittances comparable to those of FCC-ee while maintaining reasonable hardware requirements and a good packing factor that leads to a beam energy loss per turn which is as small as possible. This new arc lattice is a major improvement over the lattice presented in [3].

The arc lattice is inspired by the EBS-style achromat cell, originally developed for ultra-low-emittance light sources such as the upgrade of European Synchrotron Radiation Facility [4]. The sextupoles are placed in high-dispersion and high-beta regions to maximize chromatic correction efficiency. Phase advance between sextupoles is chosen to minimize resonance-driving terms. It is about $3\pi$ for the horizontal plane and about $\pi$ in the vertical. The main components, beta functions and horizontal dispersion can be seen in *Figure 1*.

The arc cell length is 277m long and each cell consists of 18 combined function dipoles, 12.6m long each with a small defocusing gradient of 0.5T/m at 115GeV and 4 shorter dipoles, 4.5m long. 20 quadrupoles with gradients around 21T/m (7 QFs 1.85m long, 4 QFs 1.1m long and 9 QDs 0.85m long). It also contains 6 sextupoles with gradients up to 8740T/m$^2$: 2 SFs of 0.54m long and 4 SDs, about 0.34m long.

The energy loss per turn of this lattice at 115GeV is only 5.2GeV, including an extra 4% effect coming from the two interaction regions. The corresponding bending radius is 3120m. Both of these figures represent a healthy improvement over the original submission [3] where the numbers were 5.4GeV and 2958m. Horizontal emittance is 2.3pm at 115GeV (compared to 3.8pm of [3]).

The full machine (ring circumference is 26659m) includes:

- 80 arc cells
- 8 straight sections out of which
- 2 interaction regions

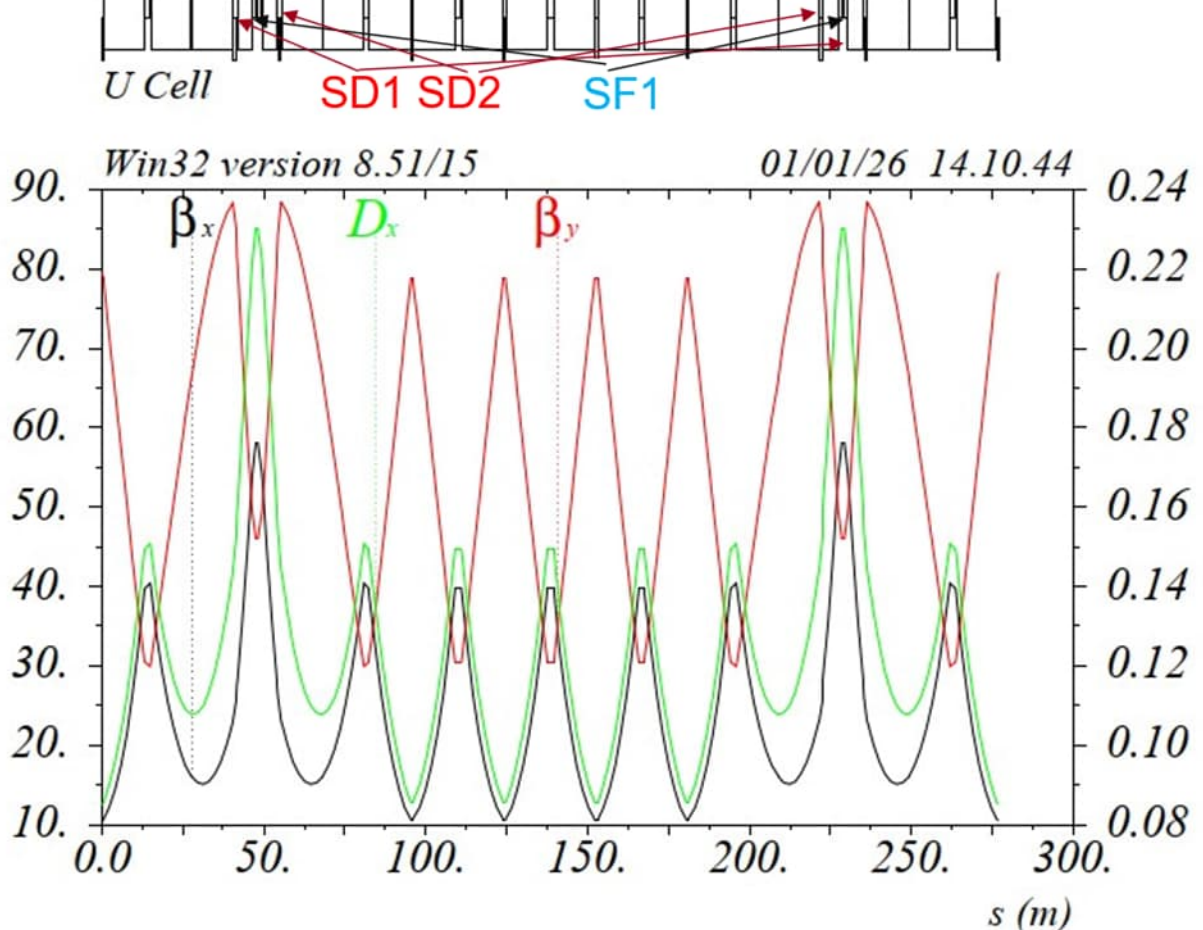

*Figure 1: the LEP3 collider arc cell. The position of sextupoles is shown.*

## STRAIGHT SECTION INSERTIONS

Eight straight sections are included in the lattice (*Figure **2***). Each straight section has a nominal length of approximately 563 m, with about 535 m usable after accounting for dispersion suppressor dipoles. These sections host: RF systems, injection systems, collimation systems, detectors and other specialized equipment, like polarization hardware. Transparency conditions are applied to preserve the periodic optical properties of the arcs. These conditions make so that the breaking of the ring periodicity is minimal, and no arc sextupole families are needed when the SSs are included in the lattice.

## FINAL FOCUS

LEP3 has two final focus (FF) systems (IP1 and PI5). The design has been developed starting from rescaling the FCC-ee local chromaticity correction design.

The system is a 5$^{th}$ order achromat in both planes (*Figure **3***). It is 1025.9mt long and has a net bend angle of 134.7mrad. The last arc octant cell per side has been replaced by the FF-Dispersion Suppressor and the FF itself. The dipoles have been optimized for best dispersion and optical functions across the Chromatic Correction Sections (CCSx/y). The dipoles in the incoming CCSy have opposite sign with respect to other dipoles, thus generating a dogleg that matches the layout constraints. The FF Left and Right magnets and drift lengths are identical and optics nearly identical, thus ensuring a high degree of symmetry.

Chromaticities due to the low-beta IP are compensated in the FF, and arc sextupoles remain unchanged. FF dipoles presently generate about 8% emittance growth. The larger contribution comes from the left_CCSx that has the stronger dipoles and a correspondingly larger curly-$\mathcal{H}$. These dipoles also increase the total energy loss/turn by about 4%. Furter optimization should mitigate these drawbacks.

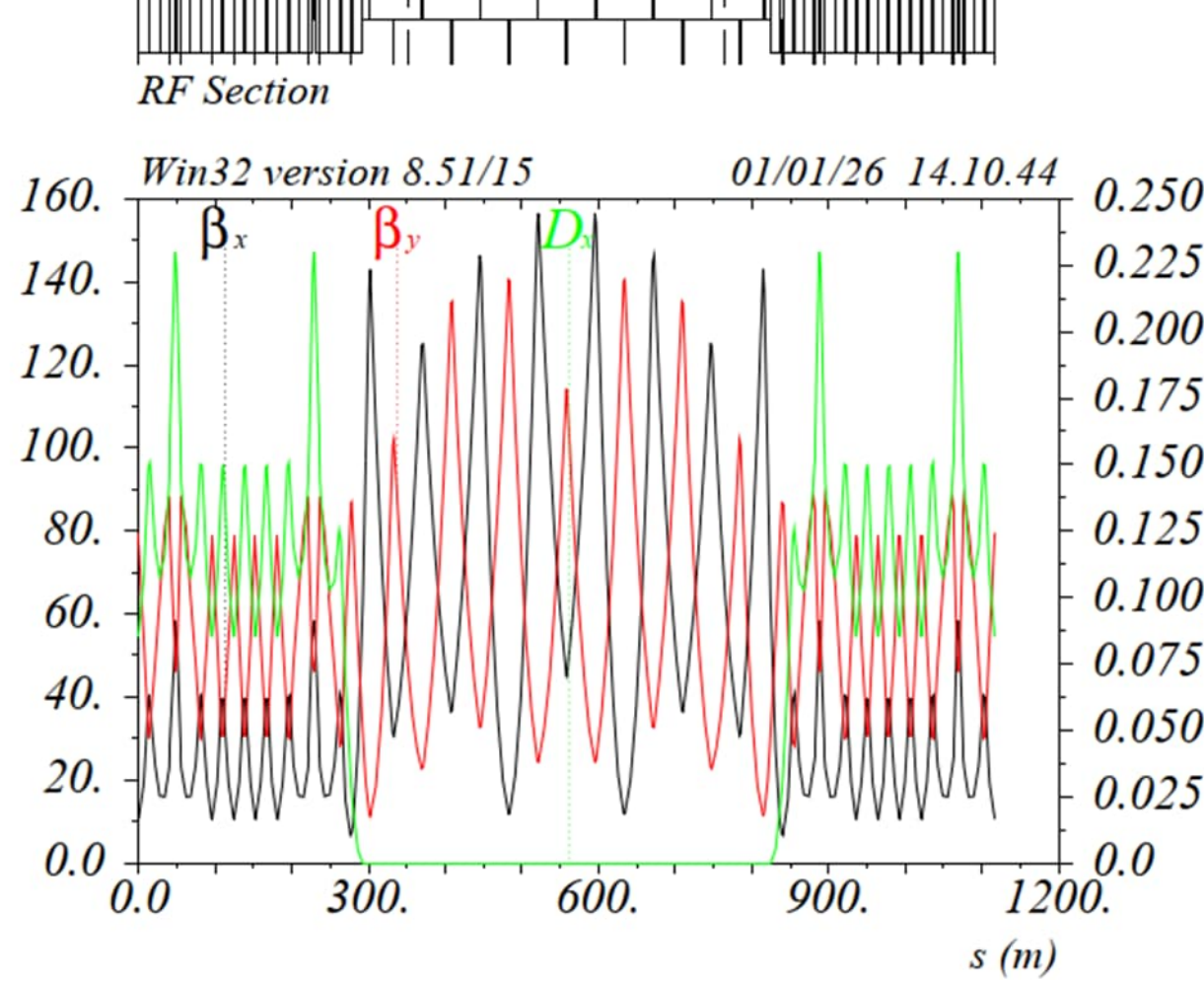

*Figure 2: a straight section insertion*

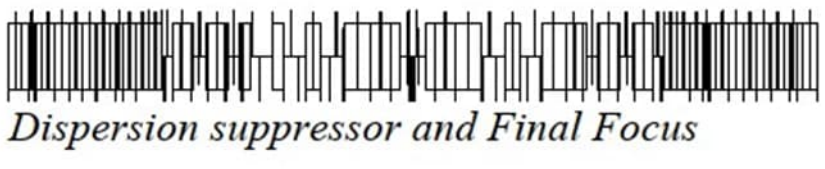

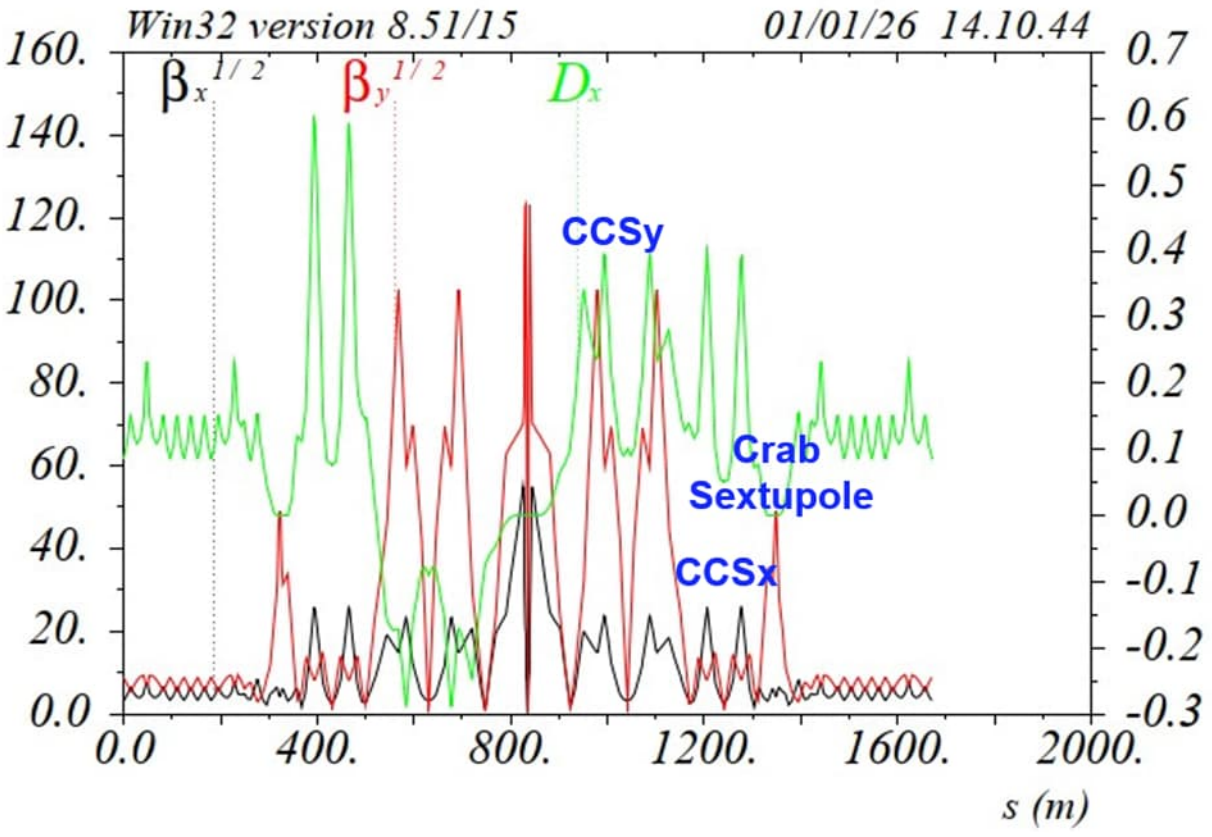

*Figure 3: the Final Focus system. The beam comes from the left. The interaction point is at 825m.*

The FF layout (*Figure* ***4***) shows a maximum beam separation of ~3.8m, which is close to being compatible with the existing tunnel (diameter 4.4m) to be verified.

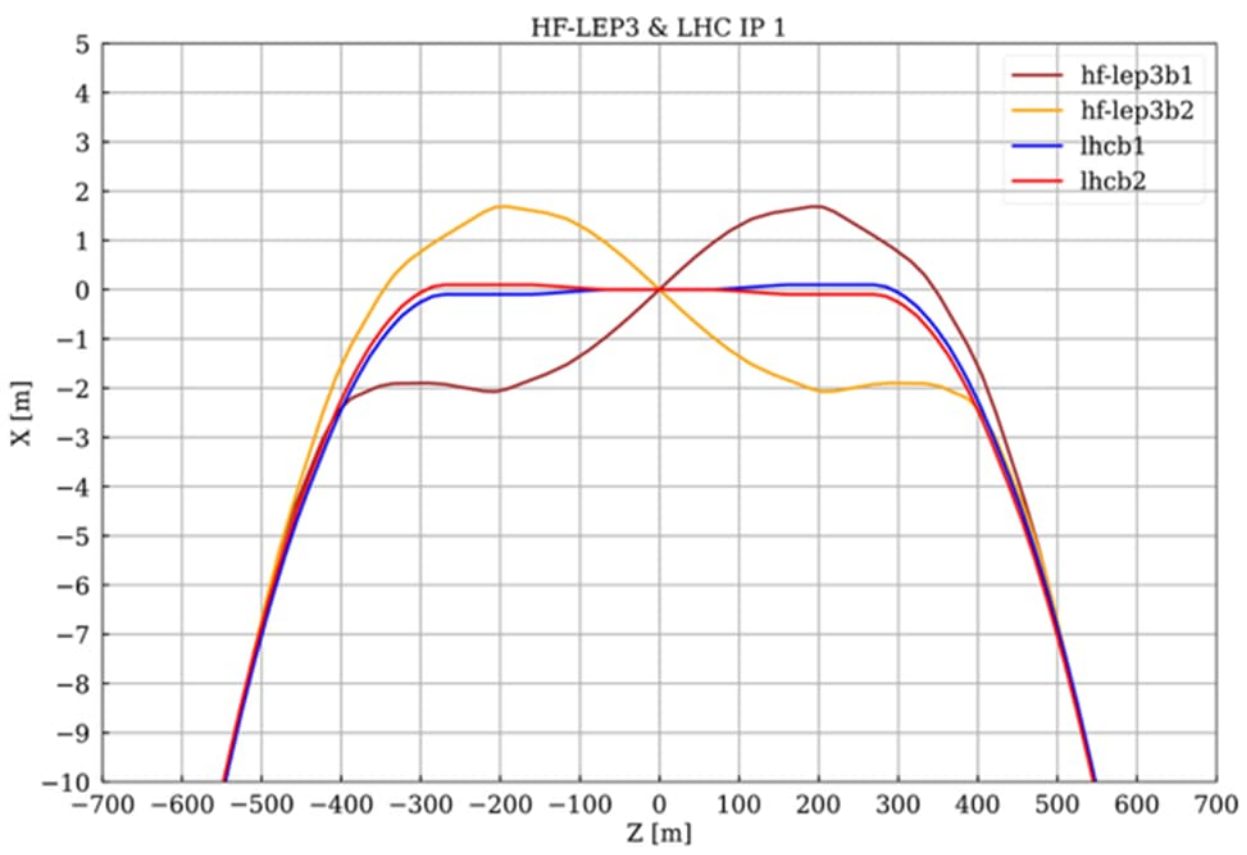

*Figure 4: final focus layout. Maximum beam separation is ~3.8m about 200m from the IP. The current LHC layout is included for comparison*

## VACUUM CHAMBER

The choice of beam pipe diameter is a critical parameter in the design of high-current electron–positron storage rings, and for LEP3 it must be optimized to achieve stable beam operation, while minimizing magnet aperture and associated hardware costs. To maintain acceptable beam stability margins, the total impedance contribution from the arcs must therefore remain within limits established by previous machines and simulation studies.

Transverse beam coupling impedance for a round vacuum chamber with radius $a$, approximately scales as:

$$Z_{\perp} \propto \frac{1}{a^3}$$

Impedance also increases linearly with the arc length and beam current and decreases linearly with the beta functions. It is also affected by longitudinal bunch length. Comparing LEP3 to FCC, LEP3 gains a factor ~4 for arc length, a factor ~3 for current and a factor ~2 for beta functions. This is a collective factor of ~24. The FCC beam pipe inner diameter is 60mm. The equivalent beam pipe diameter for LEP3, if the above scaling laws stand, would be 21mm. We propose to use a beam pipe diameter of **30mm** for LEP3 (by comparison, EBS, with 6GeV and 200mA beam current has a beam pipe diameter of 20mm). Magnet gaps should be around 38mm. This is another major deviation from [3], as magnet costs increase with the square of the magnet apertures, as does magnet power consumption.

## MAGNET DESIGN

[3] had an arc magnet power consumption, especially for the very strong arc sextupoles, that was prohibitively high for normal conducting magnets and the use of HTS magnets seemed inevitable. This design has important differences.

Dipoles are now combined function, which increases the packing factor. These are slightly more complicated than the original design and a preliminary 3D magnetic design exists (*Figure 5*).

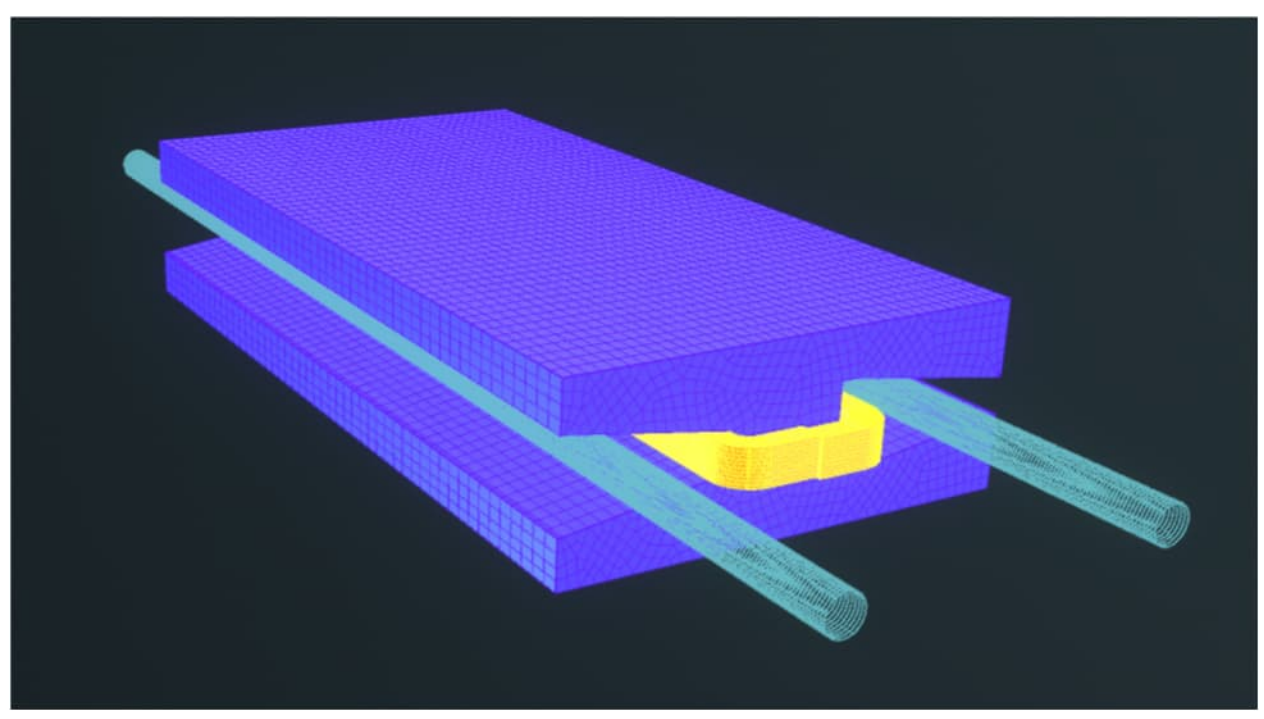

*Figure 5: combined function arc dipole*

Arc quadrupoles are now shorter and about twice as strong as the FCC-ee ones. Sextupoles are about the same strength as FCC, but total length is a lot less. The fact that both are now a factor 2 smaller, reduces their power consumption to below that of FCC for the equivalent beam energy.

Therefore, our baseline would now be to use normal conducting arc quadrupoles and sextupoles. We would like to keep the HTS option as a backup, especially since a lot of the cryogenic infrastructure needed to cool these magnets already exists in the LHC tunnel. This backup option might become the baseline in case that we would not be able to reduce the beampipe diameter to 30mm or if other capital or operational expenses can be minimized. The final focus system and crab sextupoles remain superconducting.

## RF SYSTEM

We have consolidated the RF choices of [3]. A notable difference form FCC is the use of a single frequency RF system of 800MHz. We have opted for a two-cell cavity, which naturally has no trapped modes, with a maximum operating gradient of 22.6MV/m, about 10% higher than the FCC target value. Such a cavity will provide roughly 8.4MV of acceleration per cavity. As we need 5.65GV when running at 115GeV, the number of cavities needed is around 672.

A good compromise between packing efficiency and cryomodule length is to pack 8 cavities per cryomodule. Quick calculations indicate a total length of 10.5m per cryomodule, so for 672 cavities we need 84 cryomodules of a total length of 882m.

We are also opting for both collider beams to use the same RF system. This necessitates electrostatic separators/re-combinators at every RF straight section. However, it simplifies integration issues as there is no collider beam bypass at the RF straight sections. The booster ring, sitting about a meter higher, would have to pass through, however.

We are planning to use the even points of LEP3 (points 2,4,6,8) for housing the RF. All these straight sections are 563m long and the tunnel diameter is 4.4m. They all have adjacent klystron galleries of diameter of 5.5m. There are ongoing studies to see if the available space is sufficient to house all LEP3 RF sources. The collider RF will be housed in IPs 2 and 6, leaving IPs 4 and 8 for the booster RF.

Considering that the ring is ~26.6 km long and the long straight sections ~0.5 km long, in order to avoid parasitic crossings and to ensure beam loading is as constant as possible, the filling pattern should be as follows:

- for each beam two trains of equal length and diametrically opposite
- each train is ~6.15 km long (about ¼ of the ring)
- each beam fills 46.25% of the ring
- the beams collide at IP1&IP5
- at any given time there is always a single beam in the cavities, $e^+$ and $e^-$ alternating half of the time each
- The beam loading is quasi-constant
- The large gap between the two trains is also extremely convenient for injection and beam dump kicker requirements

**Electrostatic separators**

The scheme would need a beam combination/separation system comprising of electrostatic separators and septa. We estimate that the length of the separator section would be 25 m (providing an angle of 0.4mrad) and a splitter (septum) system of length 14 m per side of the RF section. Total length is 78 m per RF section.

**Quadrupoles in the RF sections**

The long straight sections of the RF will have 12 quadrupoles interspersed, each of length of 2 m. Therefore the length taken by the EM separators, septa and quadrupoles would be 102m per RF LSS.

**Z running mode**

The peculiarities of running at the Z is high beam current (200mA to achieve a luminosity of 50E34) and the need to perform continuous resonant depolarization measurements for accurate beam energy measurement. For the latter, to achieve the smallest systematic errors, all RF should be concentrated at a single straight section.

As it turns out, the target luminosity is achieved at a SR beam power of about 25MW per beam. This means that one RF straight section housing the RF for the Z running and having 42 cryomodules and 336 cavities, would need one power coupler per cavity rated at 150kW. This represents an improvement over our previous design where new fundamental power couplers would have needed to be developed.

An obvious upgrade path would be the use of both RF straight sections for the Z running, compromising on the accuracy of the energy determination but being capable of beam power of 50MW per beam.

**H running mode**

At 115GeV the beam current is low, but the gradient requirement becomes the main focus. The 672 cavities installed, having an overall length of 882m (split in two sections) would need to operate at 22.5MV/m. This is 10% higher than the FCC target. The total length of one RF section would be 441 m for the RF, 25 m for the quadrupoles and 78 m for the separators. Total is 543m where the total length of the long straight section of the LHC tunnel is 563m.

**Booster RF**

The booster has reduced beam current and also has a lower voltage requirement of 4.2GeV. This is due to the better packing factor of the booster arcs and the judicious choice of injection points. We are also injecting off energy at -1%. For the booster we envisage to use the FCC-derived six-cell cavities as we believe they represent the cheapest choice.

The main parameters of the RF system at different energies for both the collider and booster can be seen in Table 1.

**Table 1: main RF parameters for different running modes for the collider and booster**

| Running mode | Z | WW | ZH | Booster ZH |
|---|---|---|---|---|
| Common RF for two beams | Yes | Yes | Yes | Yes |
| Total Required RF Voltage per beam (MV) | 500 | 1500 | 5700 | 4500 |
| No. of Cryomodules per beam | 43 | 86 | 86 | 60 |
| No. of LSSs per beam | 1 | 2 | 2 | 2 |
| Frequency (MHz) | 800 | 800 | 800 | 800 |
| Cells/Cavity | 2 | 2 | 2 | 6 |
| cavities/cryomodule | 8 | 8 | 8 | 4 |
| Voltage/Cavity (MV) | 1.5 | 5 | 8.4 | 22.7 |
| Power/cavity (kW) | 146 | 146 | 146 | 14 |
| length/CM | 10.5 | 10.5 | 10.5 | 9.5 |
| Beam current (mA) | 200 | 34 | 9 | 0.9 |
| Required power per beam(MW) | 25 | 50 | 50 | 5 |
| Installed voltage (MV) | 516 | 3440 | 5779 | 5448 |
| Installed power | 50 | 100 | 100 | 3 |
| Calculated length(m) | 452 | 903 | 903 | 570 |

## BOOSTER OPTICS

The booster ring optics (*Figure **6***) reproduces the collider ring optics but further exploits the advantages of combined function dipoles. This solution has been the preferred one for a large fraction of boosters in the past. The optics is virtually the same as the one implemented for ESRF-EBS, APS-U, etc. The QDs in the FODO sequence have been removed and the sequence is composed by just gradient dipoles. There are 160 QDs and QFs remaining, to provide optics flexibility. The DQ gradients are halved with respect to the QFs, further reducing the power requirements of the lattice. The dipole filling factor is 4% higher than the collider ring.

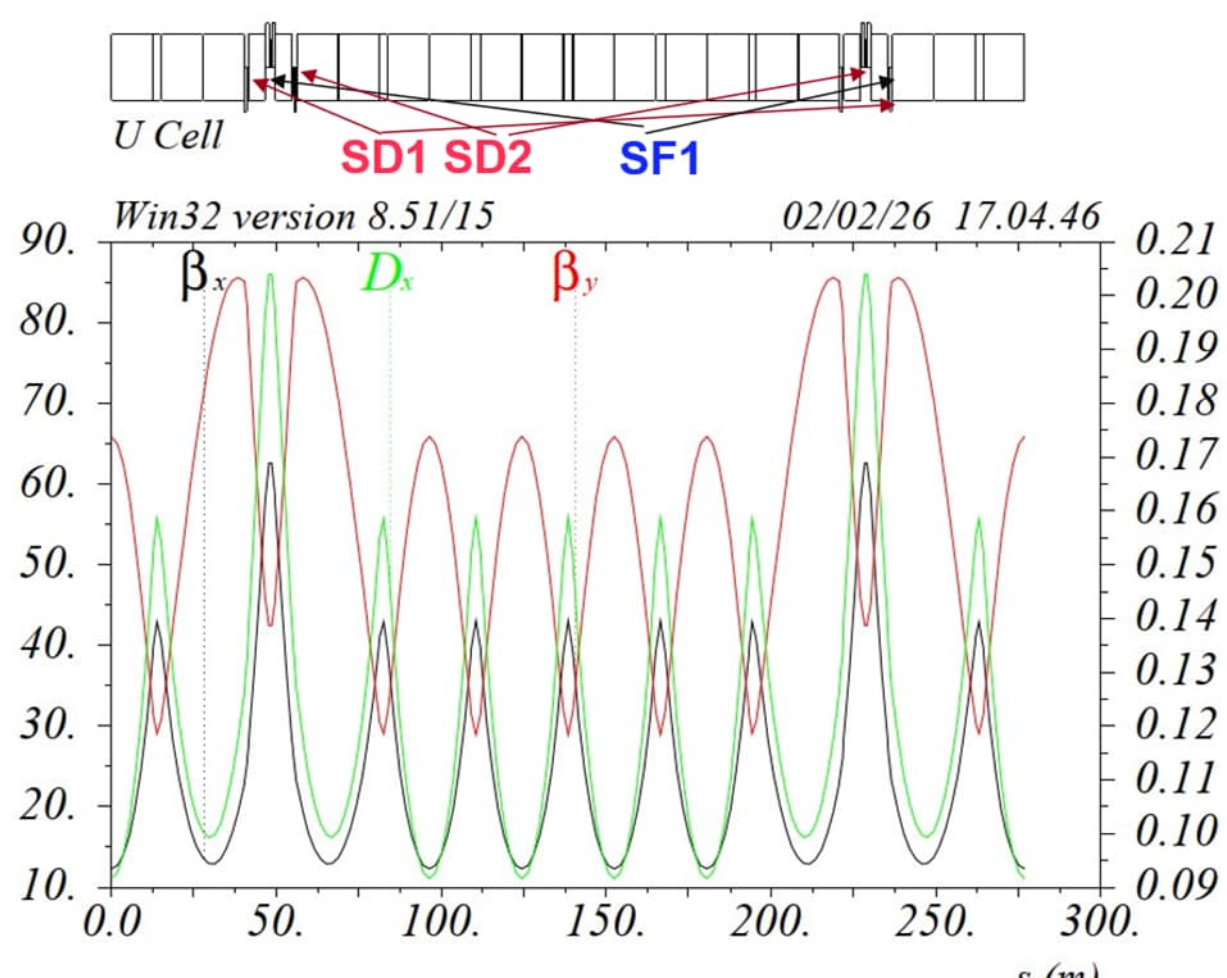

*Figure 6: Booster cell layout. The length is the same as the collider ring*

## LEP3 PARAMETER SET

We can now update our parameter set for this improved design. Since performance is greatly increased compared to [3] we have decided on a slightly different strategy: although up to now we quote performance at a fixed maximum SR power (50MW per beam) we have now decided to fix target luminosities and reduce power accordingly so as not to exceed them. These luminosities are:

Z running mode (45.6GeV): $5.0 \times 10^{35}\ cm^{-2}s^{-1}$

W running mode (80GeV): $1.0 \times 10^{35}\ cm^{-2}s^{-1}$

H running mode (115GeV): $3.0 \times 10^{34}\ cm^{-2}s^{-1}$

The Z luminosity is achieved with about 25MW per beam, whereas for the H luminosity almost the whole budget of 50MW is needed. Running at maximum power will remain an upgrade path.

The parameter list is currently not backed up by simulations. It is derived from a simple iterative model, which however predicts correctly the FCC parameters of [5].

We have managed to achieve our target luminosities using 25, 40 and 50MW of SR power per beam for the Z, W and H energies. The number of Higgs bosons produced after 6 years of operation and two experiments is now close to one million events with more than 1E12 Z bosons produced during the five-year Z campaign. The number of WW events is close to 200 million.

The horizontal emittance achieved with the new lattice is 2.3nm at 115GeV (compared to 3.8nm in [3] and 0.7nm in [5].

We have used a horizontal to vertical emittance ratio of 500 for the Z an the W, and 600 for the H running (FCC is using 340, 1000 and 700 in the feasibility study report [5].

Momentum acceptance is higher than FCC at 2% at the Z and W and 2.5% at the H. This gives acceptable but low beam lifetimes of 12, 9 and 5 minutes for the Z, W and H running.

Beam current at the Z is now 200mA and bunch population is 1.4E11 compared to 2.4E11 for FCC. To see its effect on electron cloud, we should take into account that the number of bunches compared to FCC and for the same power are down by a factor 7 with a corresponding increase to the average time between bunches. For our proposed parameter set, the number of bunches are 800 compared to 11200 in [5].

The synchrotron tune at the Z is currently 0.04. If the resonant depolarization energy measurement needs a higher synchrotron tune, we can increase the RF voltage available.

The parameter list is shown in Table 2. As the project matures, the table will be updated.

**Table 2: LEP3 updated parameter set**

| Parameter | LEP3 (ZH) | LEP3 (WW) | LEP3 (Z) |
|---|---|---|---|
| beam energy [GeV] | 115 | 80 | 45.6 |
| number of experiments | 2 | 2 | 2 |
| Circumference [m] | 26659 | 26659 | 26659 |
| crossing angle at the IP [mrad] | 30 | 30 | 30 |
| Bending radius [m] | 3122 | 3122 | 3122 |
| SR power per beam [MW] | 50 | 40.5 | 26.3 |
| beam current [mA] | 10 | 34 | 206 |
| number bunches/beam | 15 | 60 | 800 |
| bunch intensity [$10^{11}$] | 3.6 | 3.1 | 1.4 |
| SR energy loss / turn [GeV] | 5.2 | 1.2 | 0.13 |
| total RF voltage [GV] | 5.7 | 1.5 | 0.29 |
| RF frequency [MHz] | 800 | 800 | 800 |
| longitudinal damping time [turns] | 22 | 66 | 357 |
| Momentum compaction factor [$10^{-5}$] | 2.7 | 2.7 | 2.7 |
| horizontal beta* [m] | 0.20 | 0.15 | 0.10 |
| vertical beta* [mm] | 1.0 | 0.7 | 0.5 |
| horizontal emittance [nm] | 2.3 | 1.2 | 0.4 |
| vertical emittance [pm] | 3.8 | 2.3 | 0.8 |
| horizontal rms IP spot size [um] | 21 | 13 | 6 |
| vertical rms IP spot size [nm] | 62 | 40 | 20 |
| horiz. beam-beam parameter | 0.03 | 0.01 | 0.00 |
| vert. beam-beam parameter | 0.16 | 0.15 | 0.14 |
| rms bunch length SR only [mm] | 3.5 | 3.3 | 2.6 |
| rms bunch length with SR + BS [mm] | 4.7 | 6.8 | 8.8 |
| beam lifetime rad Bhabha + BS [min] | 5 | 9 | 12 |
| luminosity per IP [$10^{34}$ $cm^{-2}s^{-1}$] | 3.0 | 10.1 | 50 |
| integrated luminosity/year [ab-1] | 0.7 | 2.4 | 12.1 |
| years of operation | 6 | 3 | 5 |
| total integrated luminosity[ab-1] | 4.3 | 7.3 | 60.4 |
| total number of events [$10^6$] | 0.9 | 182 | 1.2E+06 |

## CONCLUSIONS

An updated version of the LEP3 electroweak and Higgs factory has been presented. Compared to the early design, we have improved in several areas: a low-emittance lattice has now been designed; beam pipe diameter has been reduced, and, with it, magnet cost and power consumption; the RF system has been consolidated to comfortably fit in the longitudinal space provided. Overall, the performance of the improved design is much higher than the original design at all energies.

# REFERENCES

[1] A. Blondel and F. Zimmermann, A High Luminosity e+e- Collider in the LHC tunnel to study the Higgs Boson, CERN-OPEN-2011-047 arXiv:1112.2518v1 [hep-ex], 2011.

[2] M. Koratzinos, LEP3: A possible low-cost high-luminosity Higgs factory, Proceedings of Science, IHEP-LHC-2012, 017. doi.org, 2012.

[3] Anastopoulos, C. et al., LEP3: A high-luminosity e+e- Higgs & electroweak factory in the LHC tunnel - a possible back-up to the preferred option (FCC-ee and FCC-hh) for the next accelerator for CERN, Journal of Physics G: Nuclear. Part. Physics 53 (2026) 040501.

[4] P. Raimondi et al., The Extremely Brilliant Source storage ring of the European Synchrotron Radiation Facility., Communications Physics 6, 82, 2023.

[5] FCC Collaboration, Benedikt, M., Zimmermann, F., et al. (2025), Future Circular Collider Feasibility Study Report: Volume 2 Accelerators, technical infrastructure and safety., European Physical Journal Special Topics, 234(19), 5713–6197. https://doi.org/10.1140/epjs/s11734-025-01967-4, 2025.